


\magnification\magstep1
\parskip=\medskipamount
\hsize=6.4 truein
\baselineskip=14pt
\tolerance=500


\outer\def\beginsection#1\par{\vskip0pt plus.2\vsize\penalty-300
\vskip0pt plus-.2\vsize\vskip1.2truecm\vskip\parskip
\message{#1}\leftline{\bf#1}\nobreak\smallskip\noindent}


 at 10 truept
\font\titlefont=cmbx12
\font\authorfont=cmcsc10
 at 10 truept
 at 10 truept
 4
 at 5 truept


\newdimen\itemindent \itemindent=13pt
\def\textindent#1{\parindent=\itemindent\let\par=\resetpar%
\indent\llap{#1\enspace}\ignorespaces}

\let\oldpar=\par
\def\resetpar{\oldpar\parindent=0pt\let\par=\oldpar}

\font\ninerm=cmr9 \font\ninesy=cmsy9
\font\eightrm=cmr8 \font\sixrm=cmr6
\font\eighti=cmmi8 \font\sixi=cmmi6
\font\eightsy=cmsy8 \font\sixsy=cmsy6
\font\eightbf=cmbx8 \font\sixbf=cmbx6
\font\eightit=cmti8
\def\eightpoint{\def\rm{\fam0\eightrm}
  \textfont0=\eightrm \scriptfont0=\sixrm \scriptscriptfont0=\fiverm
  \textfont1=\eighti  \scriptfont1=\sixi  \scriptscriptfont1=\fivei
  \textfont2=\eightsy \scriptfont2=\sixsy \scriptscriptfont2=\fivesy
  \textfont3=\tenex   \scriptfont3=\tenex \scriptscriptfont3=\tenex
  \textfont\itfam=\eightit  \def\it{\fam\itfam\eightit}%
  \textfont\bffam=\eightbf  \scriptfont\bffam=\sixbf
  \scriptscriptfont\bffam=\fivebf  \def\bf{\fam\bffam\eightbf}%
  \normalbaselineskip=9pt
  \setbox\strutbox=\hbox{\vrule height7pt depth2pt width0pt}%
  \let\big=\eightbig \normalbaselines\rm}
\catcode`@=11 %
\def\eightbig#1{{\hbox{$\textfont0=\ninerm\textfont2=\ninesy
  \left#1\vbox to6.5pt{}\right.\n@space$}}}
\def\vfootnote#1{\insert\footins\bgroup\eightpoint
  \interlinepenalty=\interfootnotelinepenalty
  \splittopskip=\ht\strutbox %
  \splitmaxdepth=\dp\strutbox %
  \leftskip=0pt \rightskip=0pt \spaceskip=0pt \xspaceskip=0pt
  \textindent{#1}\footstrut\futurelet\next\fo@t}
\catcode`@=12 %


\def\S{\Sigma}
\def\R{\bf R}
\def\C{\bf C}
\def\Sc{\hbox{${\cal S}$}}
\def\H{\hbox{${\cal H}$}}
\def\E{\hbox{${\cal E}$}}
\def\D{\hbox{${\cal D}$}}
\def\G{\hbox{${\cal G}$}}

\def\w{\wedge}
\def\l{\langle}
\def\r{\rangle}
\def\d{\delta}
\def\half{{1\over 2}}
\def\shalf{\hbox{${\textstyle{1\over 2}}$}}

\def\omegap{{}^{(+)}\!\omega}
\def\omegam{{}^{(-)}\!\omega}
\def\omegapm{{}^{(\pm)}\!\omega}
\def\Omegap{{}^{(+)}\!\Omega}
\def\Omegam{{}^{(-)}\!\Omega}

\def\Lambdap{{}^{(+)}\!\Lambda}

\def\Gammap{{}^{(+)}\!\Gamma}
\def\Gammam{{}^{(-)}\!\Gamma}

\def\Si{\hbox{$S_{\infty}$}}

{\titlefont
\centerline{Ashtekar Variables}
\centerline{in Classical General Relativity}}
\medskip
\centerline{\authorfont Domenico Giulini}
\vskip 1.5 truecm plus .3 truecm minus .2 truecm

\beginsection{Introduction}

The task I was given for this lecture is to present
Ashtekar's connection variables for classical General Relativity
in a pedagogical manner. My intention is to describe a possible route
towards a reformulation of General Relativity
in terms of Ashtekar's connection variables. I try
to give self-contained and hopefully painless derivations of the
crucial steps, without entering too much into those details not
directly relevant for this purpose. There exists
a comprehensive monograph on this subject (Ashtekar 1991)
which contains many applications, as well as a periodically updated
bibliography (Br\"ugmann 1993), so that the interested reader should
have no difficulties to find his/her way into the subject and the
current developments. I make no attempt to give a full account of
the current status.

In these lectures I proceed as follows: In the first chapter, I very
briefly review some elementary concepts from differential geometry, mainly
to fix notation and conventions.
Chapter two introduces the variational principle which we use to derive the
field-equations of General Relativity. Chapter three considers complex
General Relativity and shows how its field equations can be obtained from
a variational principle involving only the self-dual part of the
connection. In chapter four the (3+1)-decomposition is presented in as
much detail as seemed necessary for an audience that does not consist
entirely of canonical relativists.
It is then applied to complex General Relativity in chapter five, where
for the first time Ashtekar's connection variables are introduced. The
Hamiltonian of complex General Relativity is presented in terms of
connection variables. In chapter six the constraints that follow from the
variational principle are analyzed and their Poisson brackets are presented.
In chapter seven we discuss the reality conditions that have to be imposed
by hand to select real solutions, and briefly sketch the geometric
interpretation of the new variables. In chapter 8 we
indicate how the Hamiltonian has to be amended by surface integrals in
the case of open initial data hypersurfaces with asymptotically flat data.
It ends with a demonstration of the positivity of the mass at spatial
infinity for maximal hypersurfaces. Throughout we will make no use of
spinors.

\beginsection{Chapter 1. Some Basic Differential Geometry}

Let $M$ be a connected orientable and time orientable Lorentz 4-manifold
with topology $\S\times \R$, where $\S$ can be any connected orientable
3-manifold. Our signature convention is ``mostly plus'', $(-,+,+,+)$.
Greek indices refer to coordinate bases, latin indices to frame bases.
If they are taken from the beginning of the alphabet, i.e.,
$(\alpha,\beta,..$; $a,b,..)$, their range is $\{0,1,2,3\}$, whereas
for the middle of the alphabet, i.e.,  $(\mu,\nu,..$; $i,j,..)$, their
range is only $\{1,2,3\}$. Square brackets including
a string of n indices denote full antisymmetrization including the factor
$1/n!$. In the same way, round brackets denote symmetrization.
The Lorentz metric is denoted by $g$ (components $g_{ab}$, $g_{\alpha\beta}$).
$\eta_{ab}$ denotes the matrix $diag(-1,1,1,1)$. The components of the
curvature tensor are written $R^a_{\phantom{a}bcd}$,
$R_{ab}=R^c_{\phantom{c}acb}$ are the components of the Ricci-tensor and
$R=g^{ab}R_{ab}$ denotes the Ricci scalar.

The structure group for the real frame bundle on $M$ is $GL(4,R)$,
and $SO(1,3)$ if one restricts to orthonormal frames. We shall adopt
this restriction throughout which is equivalent to imposing an
orientation and a metric structure. We denote the
Lie algebra of $SO(1,3)$ by $so(1,3)$. Due to the assumption that
$M$ is topologically $\S\times\R$ and orientable, the frame bundle is
necessarily trivial so that we can always assume the existence of a
globally defined tetrad. This is also true if one considers the
complexified tangent bundle, which we shall need later on. We stress
that this is particular to four dimensions and will generally not hold in
dimensions three or higher than four. Let
$$\eqalignno{
&\{e_a\} \qquad\hbox{orthonormal tetrad}                      &(1.1)\cr
&\{e^a\} \qquad\hbox{orthonormal co-tetrad dual to (1.1)}\,,  &(1.2)\cr
&\hbox{so that}\quad g(e_a,e_b)=\eta_{ab}                     &(1.3)\cr
&\hbox{and}\quad e^a(e_b)=\delta^a_b                      \,. &(1.4)\cr}
$$
The volume form, $\epsilon$, on $M$ induced by $g$ is given by ($\wedge$
denotes the antisymmetric tensor product)
$$\eqalignno{
&\epsilon=e^0\wedge e^1\wedge e^2\wedge e^3=
{1\over 4!}\,\epsilon_{abcd}\,e^a\wedge e^b\wedge e^c\wedge e^d\,, &(1.5)\cr
&\hbox{where}\quad\epsilon_{abcd}=
4!\,\delta^0_{[a}\delta^1_b\delta^2_c\delta^3_{d]}\,.         &(1.6)\cr}
$$

Once we have restricted the structure group to $SO(1,3)$ we also
restrict the connections to be metric-preserving.
The connection 1-form can be represented by a globally defined
$so(1,3)$-valued 1-form, $\omega^a_b$, and the curvature by an
$so(1,3)$-valued 2-form $\Omega^a_b$. If $\nabla$ denotes the covariant
derivative, one has
$$\eqalignno{
&\nabla_{e_a} e_b=\omega^c_b(e_a)\,e_c                         &(1.7)\cr
&\eta_{ac}\omega^c_b=:\omega_{ab}=-\omega_{ba}\,,              &(1.8)\cr}
$$
and
$$\eqalignno{
&(\nabla_{e_a}\nabla_{e_b}-\nabla_{e_b}\nabla_{e_a}
-\nabla_{[e_a,e_b]})\,e_c =\Omega^d_c(e_a,e_b)\, e_d
= R^d_{\phantom{d} cab}\, e_d                                  &(1.9)\cr
&\eta_{ac}\Omega^c_b=:\Omega_{ab}=-\Omega_{ba}  \,.            &(1.10)\cr}
$$
Throughout we identify the Lie algebra with tangent
space tensors $\sigma_a^b$ such that
$\eta_{ac}\sigma^c_b=\sigma_{ab}=-\sigma_{ba}$.
The torsion is represented by an $R^4$-valued 2-form $T^a$ which
is defined through
$$
\nabla_{e_a}e_b-\nabla_{e_b}e_a-[e_a,e_b]=T^c(e_a,e_b)\,e_c\,.
\eqno{(1.11)}
$$

Let $V$ denote a vector space that carries a representation, $\rho$, of
$SO(1,3)$, and $\lambda$ a $V$-valued $n$-form on $M$ which under change
of tetrads transforms via $\rho$. Then the
exterior covariant derivative, $D$, on vector-valued forms is defined
via
$$\eqalignno{
& D\lambda:=d\lambda+\rho(\omega)\wedge \lambda \,,            &(1.12)\cr
\hbox{so that}\quad & D^2\lambda=\rho(\Omega)\wedge\lambda \,. &(1.13)\cr}
$$
Here the symbol $\rho(\omega)\wedge\lambda$ is to be understood in the
following way: The representation $\rho$ of $SO(1,3)$ on $V$ induces a
representation of its Lie algebra $so(1,3)$ on $V$, which we also denote
by $\rho$. Via this representation $\omega$ ($so(1,3)$-valued) acts on
$\lambda$ ($V$-valued), while as forms the exterior product is taken.
The exterior covariant derivatives with respect to two different
connections $\omega$ and $\omega'$ (denoted by $D^{\omega}$ and
$D^{\omega'}$) are related via
$$
D^{\omega'}\lambda=D^{\omega}\lambda+\rho(\omega'-\omega)\w \lambda \,.
\eqno{(1.14)}
$$
Applying (1.11-13) to the $R^4$-valued 1-form $\{e^a\}$ (carrying the
defining representation) one obtains the first Cartan structure equation
and the first Bianchi identity: $$\eqalignno{
& De^a=de^a+\omega^a_b\wedge e^b=T^a \,,   &(1.15)\cr
& D^2e^a=\Omega^a_b\wedge e^b =DT^a  \,.   &(1.16)\cr}
$$
By direct calculation, using the definition of the curvature, and
regarding $\Omega^a_b$ as an $so(1,3)$-valued 2-form, we obtain the
second Cartan structure equation and second Bianchi identity:
$$\eqalignno{
& d\omega^a_b+\omega^a_c\wedge\omega^c_b=
  d\omega^a_b+\shalf[\omega,\omega]^a_b=\Omega^a_b \,,      &(1.17) \cr
& D\Omega^a_b=d\Omega^a_b+[\omega,\Omega]^a_b=0        \,.  &(1.18) \cr}
$$
Here $[\quad,\quad]$ denotes the commutator if the product of
the form degrees is even, and the anticommutator if it is odd. Note also
that it would neither be algebraically correct nor meaningful
to write (1.17) as $D\omega^a_b$ (in the adjoint representation):
firstly, this expression would not have the factor 1/2 that appears in
(1.17) and, secondly, $\omega^a_b$ does not transform with any linear
transformation under change of tetrads, as would be necessary for the
covariant derivative to be meaningful. Rather, from (1.7) one easily
proves that under a change of frames: $e_a\mapsto e'_a=R^b_ae_b$,
with $SO(1,3)$-valued matrix function $R$, one has (in matrix
notation):
$$
\omega\mathop{\longmapsto}^{R}{\omega'}=R^{-1}\omega R+R^{-1} dR\,.
\eqno{(1.19)}
$$

Finally we mention the Hodge-duality map, $*$, which provides a linear
isomorphism between $n$ forms and $(4-n)$-forms at each point on $M$.
It can be defined via
$$
*(e^{a_1}\w\dots\w e^{a_n}):={1\over (4-n)!}
\epsilon^{a_1\dots a_n}_{\phantom{a_1\dots a_n}a_{n+1}\dots a_4}
e^{a_{n+1}}\w\dots\w e^{a_4}\,,
\eqno{(1.20)}
$$
and linear extension, where the indices on $\epsilon$ are raised using
$\eta^{ab}$ (the inverse matrix to $\eta_{ab}$, which has the same
entries). Since $\epsilon_{a_1a_2a_3a_4}$ is invariant under $SO(1,3)$,
the exterior covariant derivative and the $*$-operation are compatible in
the following sense:
$$
D*(e^{a_1}\w\dots\w e^{a_n})={1\over (4-n)!}
\epsilon^{a_1\dots a_n}_{\phantom{a_1\dots a_n}a_{n+1}\dots a_4}
D(e^{a_{n+1}}\w\dots\w e^{a_4})\,.
\eqno{(1.21)}
$$
Applying $*$ twice results in plus or minus the identity. In fact,
on any $n$-form $\lambda$ one has
$$
*(*\lambda)=-(-1)^{n(4-n)}\lambda\,.
\eqno{(1.22)}
$$
Analogous formulae hold in any spacetime dimensions. Here, the first
minus sign on the right hand side of (1.21) is the sign of the determinant
of $\eta_{ab}$. In four dimensions and Lorentz signature $*$ squares to
$-1$ on two forms, so that eigenforms only exist on the complexified
tangent bundle with eigenvalues $+i$ (self-dual 2-forms) and $-i$
(anti-self-dual
2-forms).
Given two $n$-forms $\lambda$ and $\sigma$ at some point in $M$:
$$\eqalignno{
& \lambda=
{1\over n!}\lambda_{a_1\dots a_n}e^{a_1}\wedge\dots\wedge e^{a_n}\,,
&(1.23)\cr
& \sigma=
{1\over n!}\sigma_{a_1\dots a_n}e^{a_1}\wedge\dots\wedge e^{a_n}\,,
&(1.24)\cr}
$$
we define their inner product, induced by $g$, via
$$
\langle \lambda,\sigma\rangle:={1\over n!}
\lambda_{a_1\dots a_n}g^{a_1b_1}\dots g^{a_nb_n}\sigma_{b_1\dots b_n}
=\lambda_{a_1\dots a_n}\sigma^{a_1\dots a_n}\,,
\eqno{(1.25)}
$$
and have
$$
\lambda\wedge * \sigma=\langle\lambda,\sigma\rangle\, \epsilon\,.
\eqno{(1.26)}
$$

\beginsection{Chapter 2. The Variational Principle}

We first show how the well known Einstein-Hilbert
action\footnote*{Throughout this article we write all actions in
their simplest form neglecting prefactors. One way to normalize the
action within pure gravity is to evaluate the energy of the
Schwarzschild solution and require it to be equal to the standard
expression. The Einstein-Hilbert action, as written down
in (2.1), gives the right energy expression in units where
$16\pi G/c^4=1$.  Note also that in the Hamiltonian formulation the
canonical momenta scale with the prefactor.}
is written in terms of the curvature 2-form $\Omega_{ab}$ and the
co-tetrads $e^a$.
$$\eqalign{
\hbox{Action}\ =
S:= & \int\Omega_{ab}\w * (e^a\w e^b)                               \cr
  = & \shalf\int R_{abcd}(e^c\w e^d)\w * (e^a\w e^b)                \cr
  = & \shalf\int R_{abcd}\l e^c\w e^d\,,\,e^a\w e^b\r\,\epsilon       \cr
  = & \shalf\int R_{abcd}(\eta^{ac}\eta^{bd}-\eta^{ad}\eta^{bc})
                                                        \,\epsilon  \cr
  = & \int R\,\epsilon\ =\ \hbox{Einstein-Hilbert Action.}           \cr}
\eqno{(2.1)}
$$

We now take the co-tetrads $e^a$ and connection 1-forms $\omega^a_b$
as independent variables with respect to which we vary the action $S$.
We stress that since we restricted the connection 1-form to be
$so(1,3)$-valued we have put in metricity of the connection by hand.
Varying the curvature with respect to the connection 1-form yields:
$$
\d \Omega_{ab}=d(\d\omega_{ab}) + [\omega,\d\omega]_{ab}=D(\d\omega_{ab})\,,
\eqno{(2.2)}
$$
so that the variation of $S$ with respect to $\omega$, denoted by
$\d^{\omega}$, is given by
$$
\d^{\omega}S =
\shalf\int \d\omega_{ab}\w \epsilon^{ab}_{\phantom{ab}cd}\, D(e^c\w e^d)\,,
\eqno{(2.3)}
$$
where we have performed an integration by parts and used the fact that the
variations vanish at the boundary of the integration domain.
The requirement of stationarity is thus equivalent to:
$$
\d^{\omega}S=0\Leftrightarrow D(e^a\w e^b)=0=T^a\w e^b-e^a\w T^b\,.
\eqno{(2.4)}
$$
It is not difficult to show directly from the rightmost expression that
this implies vanishing torsion. But, keeping an eye towards later
generalizations,
we shall proceed differently. We write $\omega=\Gamma+\Lambda$
where $\Gamma$ is the Levi-Civita connection, i.e., the unique metric,
torsion free connection. If we denote the exterior covariant derivative
with respect to $\Gamma$ ($\omega$) by $D^{\Gamma}$ ($D^{\omega}$), we have,
using (1.13) and the vanishing torsion of $\Gamma$,
$$\eqalign{
D^{\omega}(e^a\w e^b)= & D^{\Gamma}(e^a\w e^b)+
                       \Lambda^a_c\w e^c\w e^b+\Lambda^b_c\w e^a\w e^c \cr
                     = & (\Lambda^a_{[cd]}\d^b_e+\Lambda^b_{[ce]}\d^a_d)\,\,
                          e^c\w e^d\w e^e \,,\cr}
\eqno{(2.5)}
$$
where we used the expansion $\Lambda^a_b=\Lambda^a_{cb}e^c$.
This last expression vanishes precisely if the cyclic sum in $c-d-e$ of
the coefficient does:
$$
\Lambda^a_{[cd]}\d^b_e + \Lambda^b_{[ce]}\d^a_d  +
\Lambda^a_{[de]}\d^b_c + \Lambda^b_{[dc]}\d^a_e  +
\Lambda^a_{[ec]}\d^b_d + \Lambda^b_{[ed]}\d^a_c\ = 0 \,.
\eqno{(2.6)}
$$
Contraction on $b$ and $e$ yields:
$$
\Lambda^a_{[cd]} + \Lambda^b_{[cb]}\d^a_d - \Lambda^b_{[db]}\d^a_c = 0\,.
\eqno{(2.7)}
$$
Another contraction on $a$ and $d$ implies $\Lambda^a_{[ca]}=0$, which,
after reinsertion into (2.7), implies $\Lambda^a_{[cd]}=0$, i.e.
symmetry of $\Lambda^a_{bc}$ in the lower index pair.
In its purely covariant form: $ \eta_{bd}\Lambda^d_{ac}=:\Lambda_{abc}$,
we obtain a tensor which is symmetric in the first and third, and
antisymmetric in the second and third slot, where the antisymmetry
follows from the metricity of $\omega$ and $\Gamma$. But such a tensor is
necessarily zero as three consecutive applications of interchangements
of the first with third followed by second with third index show.
Note that $\Lambda^a_{[bc]}=0$ is just the condition for $\omega$ and
$\Gamma$ having the same torsion, so that the last arguments may be
seen as a uniqueness proof for metric, torsion free connections.
As result of the variation $\d^{\omega}$ we thus obtain
$$
\d^{\omega}S=0\Longleftrightarrow \omega=\Gamma=
\hbox{Levi-Civita connection.}
\eqno{(2.8)}
$$
For the Levi-Civita connection $\Gamma$ we can invert the first Cartan
structure equation (1.15) and express it purely in terms of the co-tetrad
$e^a$:
$$
\Gamma^{ab}=-*\left[(*de^a)\w e^b-(*de^b)\w e^a+\shalf (e^a\w e^b)\w *
(de^c\w e^d)\,\eta_{cd}\right]\,.
\eqno{(2.9)}
$$

Next, we take the variation with respect to the co-tetrad $e^a$,
which we denote by $\d^e$. The variation of the integrand yields:
$$\eqalign{
\d^e\,\left(\Omega_{ab}\w * (e^a\w e^b)\right)
&= \left(\epsilon^{ab}_{\phantom{ab}cd}\Omega_{ab}\w e^c
\right)\w\d e^d \cr
&= \shalf\left(\epsilon^{ab}_{\phantom{ab}cd}\,R_{abgh}\, e^g\w e^h\w e^c
\right)\w\d e^d \cr
&= \shalf\left(\epsilon_{abcd}\,R^{ab}_{\phantom{ab}gh}\,
\epsilon^{ghce}\,\eta_{ef} *e^f\right)\w\d e^d \cr
&= -\shalf\left(3!\,\d^g_{[a}\d^h_b\d^e_{d]}\, R^{ab}_{\phantom{ab}gh}\,
\eta_{ef} *e^f\right)\w\d e^d \cr
&= 2\left(R_{fd}-\shalf\eta_{fd}R\right)*e^f\w\d e^d \,,\cr}
\eqno{(2.10)}
$$
so that
$$\eqalignno{
\d^e S= 0 & \Longleftrightarrow {\d S\over \d e^b}=
2\left(R_{ab}-\shalf\eta_{ab}R\right)*e^a=0                         &(2.11)\cr
          & \Longleftrightarrow G_{ab}=R_{ab}-\shalf\eta_{ab}R=0\,. &(2.12)\cr}
$$
If we finally insert into this variation formula the result of the first
variation, $\omega=\Gamma$, where $\Gamma$ is to be understood as a function
of the co-tetrad according to (2.9), then we obtain equations for the
co-tetrads which are equivalent to Einstein's vacuum equations.

\beginsection{Chapter 3. Complex GR and Self-Dual Variational Principle}

We now extend the differential-geometric framework and consider the
{\it complexified} tensor bundle
$$
T_{\C}=\bigoplus_{r,s}T^r_s(M)\otimes \C \,,
\eqno{(3.1)}
$$
over the {\it real} manifold $M$. Here $T^r_s(M)$ denotes the (real)
tensor bundle of $r$-fold contravariant and $s$-fold covariant tensors
over $M$. The structure group is now $SO(1,3)_{\C}\cong SO(4)_{\C}$.
The fields $e^a$, $\omega^a_b$, $\Omega^a_b$, $T^a$ and $\eta_{ab}$
extend to sections in the complexified bundle. Everything we have
said so far is still valid in the complex case. In particular, the
variational principle and the resulting equations of motion transcribe
verbatim with all quantities now being complex. The theory so obtained
may be called complex General Relativity. Since the complex tensor bundle
was obtained by complexification of a real tensor bundle, there is a natural
reality structure. In other words, we know how to complex conjugate, namely
by complex conjugation of the components with respect to a real basis which
is naturally given to us since we work over a real manifold.
We can thus speak of real sections in the complex tensor bundle.

On the space of complex 2-forms we can now consider eigenforms to the
$*$-operator (remember, on 2-forms: $**=-$ identity)
$$\eqalignno{
& *\lambda= i\lambda \quad\hbox{self-dual}\,,      &(3.2a)\cr
& *\lambda=-i\lambda \quad\hbox{anti-self-dual}\,. &(3.2b)\cr}
$$
Similarly, in the complexified Lie algebra $so(1,3)_{\C}$, we can define a
dualization operation (denoted by $\star$):
$$
\star\sigma_{ab}:=\shalf\epsilon_{ab}^{\phantom{ab}cd}\sigma_{cd}\,
\eqno{(3.3)}
$$
that also squares to minus the identity and thus decomposes the
complexified Lie algebra into self- and anti-self-dual part:
$$\eqalignno{
& so(1,3)_{\C}^{(+)}=\{\sigma\in so(1,3)\ /\ \star\sigma= i\sigma\}
\quad\hbox{self-dual}\,,       &(3.4a)\cr
& so(1,3)_{\C}^{(-)}=\{\sigma\in so(1,3)\ /\ \star\sigma=-i\sigma\}
\quad\hbox{anti-self-dual}\,.  &(3.4b)\cr}
$$
Using the standard formula for multiplying two $\epsilon$-tensors
one easily shows the following relations for commutators:
$$\eqalignno{
& \star[\sigma,\star\tau]=-[\sigma,\tau]=\star[\star\sigma,\tau]\,,&(3.5)\cr
\hbox{hence}\quad
& \star[\sigma,\tau]=[\star\sigma,\tau]=[\sigma,\star\tau]       &(3.6)\cr
\hbox{and}\quad
& [\star\sigma,\star\tau]=-[\sigma,\tau]  \,.                    &(3.7)\cr}
$$
If we denote the projectors onto the self-dual and anti-self-dual
parts of $so(1,3)_{\C}$ by $P^{(+)}$ and $P^{(-)}$:
$$
P^{(\pm)}=\shalf(1\mp i\star)\,,
\eqno{(3.8)}
$$
then (3.5-7) imply
$$
P^{(\pm)}[\sigma,\tau]=[P^{(\pm)}\sigma,\tau]=[\sigma,P^{(\pm)}\tau]=
                       [P^{(\pm)}\sigma,P^{(\pm)}\tau]\,.
\eqno{(3.9)}
$$
This tells us that $so(1,3)_{\C}^{(\pm)}$ form ideals in $so(1,3)_{\C}$
with projection homomorphisms $P^{(\pm)}$. We thus have
a decomposition of the Lie algebra into two ideals:
$$
so(1,3)_{\C}=so(1,3)_{\C}^{(+)}\oplus so(1,3)_{\C}^{(-)}\,.
\eqno{(3.10)}
$$
$so(1,3)$ and $so(1,3)_{\C}^{(\pm)}$ are simple Lie algebras whereas
$so(1,3)_{\C}$ is only semi-simple.

We can now decompose $\omega$ and $\Omega$ according to (3.9):
$$\eqalignno{
& \omega=\omegap + \omegam = P^{(+)}\omega + P^{(-)}\omega\,,  &(3.11)\cr
& \Omega=\Omegap + \Omegam = P^{(+)}\Omega + P^{(-)}\Omega
        = \Omega(\omegap) + \Omega(\omegam)                 \,, &(3.12)\cr}
$$
where $\Omega(\omegapm)$ is the curvature built from $\omegapm$ according
to (1.17). It should be noted that $\omegapm$ are not connections on the
original bundle with structure group $SO(1,3)_{\C}$ (e.g. a general gauge
transformation in $SO(1,3)_{\C}$ will not leave it
$so(1,3)_{\C}^{(\pm)}$-valued), but on associated bundles whose structure
groups are given by the subgroups in $SO(1,3)_{\C}$ that correspond to the
Lie sub-algebras $so(1,3)_{\C}^{(\pm)}$.
Noting that $\Omega_{ab}\w *(e^a\w e^b)=\star\Omega_{ab}\w (e^a\w e^b)$,
we can decompose the action of complex General Relativity:
$$\eqalign{
S_{\C}=S^{(+)}_{\C}+S^{(-)}_{\C}
=  \int\Omegap_{ab}\w * (e^a\w e^b) +
   \int\Omegam_{ab}\w * (e^a\w e^b)     \cr
= i\int\Omegap_{ab}\w   (e^a\w e^b)
 -i\int\Omegam_{ab}\w   (e^a\w e^b) \,. \cr}
\eqno{(3.13)}
$$
The variation of the actions $S_{\C}$ and $S^{(\pm)}_{\C}$ with respect to
$\omega$, respectively $\omegapm$, will in each case result in a
determination of these connections in terms of the co-tetrad fields.
The variation with respect to the co-tetrad fields, evaluated
at the particular value of the connections just determined, then result in
equations for the co-tetrad fields only. Their solutions  will then determine
connections $\omega$ (if $S_{\C}$ is varied) or $\omegapm$
(if $S^{(\pm)}_{\C}$ is varied) in the corresponding bundles.
The crucial property of these different variational principles is that they
result in the {\it same} equations for the co-tetrads.
Let us summarize this in the following

\goodbreak
\proclaim Theorem.
The stationary points of $S_{\C}$ and $S^{(\pm)}_{\C}$ lie over the same
co-tetrad fields. Both actions assume the value zero at their stationary
points.

\nobreak
\noindent
{\bf Proof.} The variation of $S_{\C}$ proceeds identically to
the variation of the real action $S$. With $S^{(\pm)}_{\C}$ we also
proceed as in the real case. We shall only deal with $S^{(+)}_{\C}$
since the anti-self-dual case is entirely analogous.
First varying $\omegap$ we obtain
(where we denote the exterior covariant derivative with respect to
$\omegapm$ by ${}^{(\pm)}D$)
$$\eqalign{
\d^{\omegap}S_{\C}^{(+)}
= &  \int {}^{(+)}\!D(\d\omegap_{ab})\w * (e^a\w e^b) \cr
= & i\int \d\omegap_{ab}\w {}^{(+)}\!D (e^a\w e^b)     \cr
= & 0 \Leftrightarrow {}^{(+)}\!D(e^a\w e^b)^{(+)}=0     \cr}
\eqno{(3.14)}
$$
where $(e^a\w e^b)^{(+)}$ denotes the self-dual part of $(e^a\w e^b)$
considered as a Lie algebra valued 2-form.
We set
$\omegap=\Gammap+\Lambdap=\Gamma - \Gammam+\Lambdap$ and have, using that
$\Gamma$ has vanishing torsion and that the anti-self-dual $\Gammam^{ab}$
commutes with the self-dual $(e^a\w e^b)^{(+)}$, analogous to formula (2.5),
$$\eqalign{
\d^{\omegap}S^{(+)}=0\Leftrightarrow
  &  \Lambdap^a_c \w (e^c\w e^b)^{(+)}
  +  \Lambdap^b_c \w (e^a\w e^c)^{(+)} \cr
= &  \Lambdap^a_c \w (e^c\w e^b)
  +  \Lambdap^b_c \w (e^a\w e^c)       \cr
= &  (\Lambdap^a_{[cd]}\d^b_e+\Lambdap^b_{[ce]}\d^a_d)\,e^c\w e^d\w e^e\,,\cr}
\eqno{(3.15)}
$$
where we used the expansion $\Lambdap^a_b=\Lambdap^a_{cb}e^c$. We can now
conclude exactly as in (2.5) and (2.6), with $\Lambdap$ replacing $\Lambda$,
and obtain the analogous formula to (2.7). This leads to
$$\eqalign{
\d^{\omegap}S^{(+)}=0
\Longleftrightarrow \omegap=\Gammap=\, & \hbox{self-dual part of the}\cr
                                     & \hbox{Levi-Civita connection.}\cr}
\eqno{(3.16)}
$$
As in the real case, $\omegap$ is now determined as function of the
co-tetrad by taking the self dual part of (2.9).

Next we perform the variation with respect to the co-tetrad.
The variational derivative is easily obtained:
$$
{\d S^{(+)}_{\C}\over \d e^b} = 2i\,\Omegap_{ab}\w e^a\,.
\eqno{(3.17)}
$$
Evaluated at $\omegap=\Gammap$, this is equal to
$$\eqalignno{
{\d S^{(+)}_{\C}\over \d e^b}\Big\vert_{\omegap=\Gammap}
& = i\,\left(\Omega_{ab}\w e^a - i\shalf
    \epsilon^{cd}_{\phantom{cd}ab}\Omega_{cd}\w e^a\right)    &(3.18)\cr
& = \shalf\epsilon^{cd}_{\phantom{cd}ab}\Omega_{cd}\w e^a     &(3.19)\cr
& = \left(R_{ab}-\shalf\eta_{ab}R\right)*e^a                  &(3.20)\cr
& = \half{\d S_{\C}\over \d e^b}\Big\vert_{\omega = \Gamma} \,,    &(3.21)\cr}
$$
where from (3.18) to (3.19) we have used the first Bianchi identity
(1.16) and the step from (3.19) to (3.20) is entirely analogous to
the calculation leading to (2.10). Equation (3.20) then follows from
comparison with (2.11), simply transcribed for the complex case.
Finally, we observe that at the stationary
points the actions $S_{\C}^{(\pm)}$ and $S_{\C}$ assume the value zero.
Note also the factor $\shalf$ in (3.21), which is just the $\shalf$ from
the projection operation (formula (3.8)) onto the self-dual part.
It implies that the properly normalized self-dual action should have twice
the prefactor of the action $S_{\C}$.
This concludes our proof of the theorem $\bullet$.

\beginsection{Chapter 4. The 3+1-Decomposition}

Let $M\cong \R\times\S$ be foliated by the images of a one-parameter
family of embeddings, $\E_t\,:\,\S\rightarrow M$, and let $\S_t$ denote
the image of $\S$ under $\E_t$ in $M$. We have the following vector field
over $\E_t$:
$$
{\partial\over\partial t}:={d\over dt}\E_t\,\;\quad
p\mapsto{d\over dt}\E_t(p)\in T_{\E_t(p)}(M)\,,
\eqno{(4.1)}
$$
whose spacetime interpretation is to generate the motion of $\S$
through $M$. We decompose it with respect to an orthonormal tetrad
$\{e_{\perp},e_k\}$ which is adapted to the foliation. The dual co-tetrad
is $\{e^{\perp},e^k\}$. Note that the $e_k$'s are chosen tangential to
$\S$, which does not imply the $e^k$'s to be tangent as well (i.e. a section
in the cotangent bundle of $\S$). Rather, the dual basis to $\{e_k\}$,
intrinsic to $\S$, is given by the projections of the $e^k$'s parallel
to $\S$ and will instead be called $\{\theta^k\}$. We will not use it
until the last section. We now have:
$$
{\partial\over \partial t} =Ne_{\perp}+N^ke_k  \,,
\eqno{(4.2)}
$$
where $N$ and $N^ke_k$ are the so-called lapse function and shift vector
field respectively. Let further $\{x^{\mu}\}$ be a coordinate system on
$\S$ so that $\{t,x^{\mu}\}$ is an adapted coordinate system on $M$ and
where $\partial/\partial t$ is given by (4.2). We shall from now on write
a $\perp$ for the zeroth frame index and $t$ for the zeroth coordinate
to distinguish the two and to remind on the special choice made.

The relation between the spatial coordinate and frame bases is given by
$$
{\partial\over\partial x^{\mu}}=e_{\mu}^k e_k\,,\quad
e_k=e_k^{\mu}{\partial\over\partial x^{\mu}}\,.
\eqno{(4.3)}
$$
where $\{e^k_{\mu}\}$ and $\{e^{\mu}_k\}$ are relative inverse matrices.
We have ($N^{\mu}$ denote the components of the shift vector with respect
to the coordinate basis)
$$\eqalignno{
   \pmatrix{\partial_t\cr\partial_{\mu}}
=& \pmatrix{N & N^k      \cr 0 & e^k_{\mu}\cr}
   \pmatrix{e_{\perp}\cr e_k \cr}                      &(4.4)\cr
   \pmatrix{e_{\perp}\cr e_k \cr}
=& \pmatrix{1/N & -N^{\mu}/N \cr 0 & e_k^{\mu}\cr}
   \pmatrix{\partial_t \cr \partial_{\mu}     \cr}     &(4.5)\cr}
$$
and
$$\eqalignno{
\left(dt\,,\,dx^{\mu}\right)
= & \left(e^{\perp}\,,\,e^k\right)
    \pmatrix{1/N & -N^{\mu}/N \cr 0 & e_k^{\mu}\cr}      &(4.6)\cr
\left(e^{\perp}\,,\,e^k\right)
= & \left(dt\,,\,dx^{\mu}\right)
    \pmatrix{N & N^k \cr 0 & e_{\mu}^k \cr}\,.           &(4.7)\cr}
$$
The four-dimensional Lorentz metric then decomposes as follows:
$$\eqalign{
ds^2 &= \eta_{ab}\ e^a\otimes e^b =
        g_{\alpha\beta}\ dx^{\alpha}\otimes dx^{\beta}               \cr
     &= -e^{\perp}\otimes e^{\perp} + \sum_{k=1}^3 e^k\otimes e^k    \cr
     &= -N^2 dt\otimes dt + \sum_{k=1}^3 (N^k dt+e^k_{\mu} dx^{\mu})\otimes
                                        (N^k dt+e^k_{\nu}dx^{\nu})\,,\cr}
\eqno{(4.8)}
$$
where the coordinate components of the spatial part are given by
$$\eqalignno{
& h_{\mu\nu}= \sum_{k=1}^3 e^k\otimes e^k (\partial_{\mu},\partial_{\nu})
             = \sum_{k=1}^3 e^k_{\mu} e^k_{\nu}                &(4.9)\cr
\hbox{and inverse}\quad
& h^{\mu\nu}= \sum_{k=1}^3 e_k^{\mu} e_k^{\nu}\,,              &(4.10)\cr
\hbox{so that}\quad  & e_k^{\mu}=h^{\mu\nu}\d_{kl}e_{\nu}^l
\quad\hbox{and}\quad e_{\mu}^k=h_{\mu\nu}\d^{kl}e_l^{\nu}\,.   &(4.11)\cr}
$$
In words: $e^k_{\mu}$ is obtained from $e_l^{\nu}$ by raising the
(latin) frame index with the Kronecker delta $\d^{kl}$ ($=\d_{kl}$)
and lowering the (greek) coordinate index with $h_{\mu\nu}$.
Using (4.10-11) and the definition $N_{\mu}:=h_{\mu\nu}N^{\nu}$, we can
write the Lorentz metric and its ``inverse'' in the following form:
$$\eqalignno{
& g_{\alpha\beta}\,dx^{\alpha}\otimes dx^{\beta}
  = - e_{\perp}\otimes e_{\perp} + \sum_{k=1}^3 e_k\otimes e_k    & \cr
& = - (N^2-N^{\mu}N_{\mu})\,dt\otimes dt
    + N_{\mu}\,(dt\otimes dx^{\mu}+dx^{\mu}\otimes dt)
    + h_{\mu\nu}\,dx^{\mu}\otimes dx^{\nu}               & (4.12)\cr
& g^{\alpha\beta}\,\partial_{\alpha}\otimes\partial_{\beta}
  = - e^{\perp}\otimes e^{\perp} + \sum_{k=1}^3 e^k\otimes e^k   &\cr
&= - {1\over N^2}\partial_t\otimes\partial_t
   + {N^{\mu}\over N^2}\left(\partial_t\otimes \partial_{\mu}+
      \partial_{\mu}\otimes\partial_t\right)
   +\left(h^{\mu\nu}- {N^{\mu}N^{\nu}\over N^2}\right)\,
     \partial_{\mu}\otimes\partial_{\nu}    \,,              &(4.13)\cr}
$$
or simply:
$$\eqalignno{
& \{g_{\alpha\beta}\}
  =\pmatrix{-N^2+N^{\mu}N_{\mu} & N_{\mu}     \cr
             N_{\mu} & h_{\mu\nu}             \cr}          &(4.14)\cr
& \{g^{\alpha\beta}\}
  =\pmatrix{-1/N^2 & N^{\mu}/N^2               \cr
  N^{\mu}/N^2 & h^{\mu\nu}-N^{\mu}N^{\nu}/N^2 \cr} \,.      &(4.15)\cr}
$$
The square-root of the determinant of $\{g_{\alpha\beta}\}$,
denoted by $\sqrt{g}$, is most easily obtained from (4.7,9) and the
definition of the volume 4-form (1.5) (the square-root of the
determinant of $\{h_{\mu\nu}\}$ is called $\sqrt{h}$):
$$\eqalign{
\epsilon = & e^{\perp}\w e^1\w e^2\w e^3
         =   N \hbox{det}\{e_{\mu}^k\}\ dt\w dx^1\w dx^2\w dx^3\cr
         = & N\sqrt{h}\ dt\w dx^1\w dx^2\w dx^3                \cr
         = &  \sqrt{g}\ dt\w dx^1\w dx^2\w dx^3 \,,\quad
               \hbox{hence}\quad \sqrt{g}=N\sqrt{h}          \cr}
\eqno{(4.16)}
$$
Note that
$$
\sqrt{h}=\hbox{det}\{e_{\mu}^k\}=1/\hbox{det}\{e^{\mu}_k\}\,.
\eqno{(4.17)}
$$

If we were to follow the standard route and put the action principle
(2.1) into Hamiltonian form, we would decompose the co-tetrad
variables according to (4.7), i.e., we would use the variables
$$
            \left(e^k_{\mu}\,,\,N^{\mu}\,,\, N\right)\,.
\eqno{(4.18)}
$$
A first step in arriving at an algebraically simpler form of
the Hamiltonian action integral is to take the frame rather
than co-frame components as variables and
redefine these variables by partially densitizing them
(i.e. multiplying by powers of $\sqrt{h}$).
The new variables are:
$$\eqalignno{
& E_k^{\mu}:= \sqrt{h}\,e^{\mu}_k =
   {1\over\hbox{det}\{e^{\mu}_k\}} e^{\mu}_k   &(4.19)\cr
& N'^{\mu}:=  N^{\mu} \,,                      &(4.20)\cr
& N':=       {1\over\sqrt{h}}\,
             N=\hbox{det}\{e_k^{\mu}\}\,N \,.  &(4.21)\cr}
$$
We have
$$
\{dx^{\alpha}(e_a)\}=\pmatrix{1/N & 0\cr -N^{\mu}/N & e^{\mu}_k \cr}
= {1\over\sqrt{h}}\pmatrix{1/N'& 0 \cr -N'^{\mu}/N' & E_k^{\mu} \cr}
=:{1\over\sqrt{h}}\,E_a^{\alpha}
\eqno{(4.22)}
$$
and
$$\eqalignno{
\sqrt{g} =& N\sqrt{h} = N'h                                      &(4.23)\cr
h =& [\hbox{det}\{e_k^{\mu}\}]^{-2} = \hbox{det}\{E_k^{\mu}\}\,. & (4.24)\cr}
$$

Now we have all the necessary formulae to write the action
$$
S=\int\Omega_{ab}\w * (e^a\w e^b)
\eqno{(4.25)}
$$
in 3+1-decomposed form using the densitized variables just introduced.
Since from now on the old variables will never appear, we remove the
dashes from these variables. But remember: from now on $E_k^{\mu}$
are the components of a triad of density weight-one, $N^{\mu}$ is a
spatial vector field and $N$ is a spatial scalar of density weight
minus one. The action integrand then decomposes as follows:
$$\eqalignno{
& \Omega_{ab}\w * (e^a\w e^b)
   = \shalf R_{ab\alpha\beta}\,dx^{\alpha}\w dx^{\beta}
                                      \w * (e^a\w e^b)         &\cr
& = \shalf R_{ab\alpha\beta}\,\l dx^{\alpha}\w dx^{\beta}\,,
                 \,e^a\w e^b\r\epsilon                         &\cr
& = \shalf R_{ab\alpha\beta}\,E_c^{\alpha}E_d^{\beta}\,
          \l e^c\w e^d\,,\,e^a\w e^b\r\,{1\over h}\epsilon     &\cr
& = R^{cd}_{\phantom{cd}\alpha\beta} E_c^{\alpha} E_d^{\beta}\,
           N\, dtd^3x                                          &\cr
& = \left( N\, R^{kl}_{\phantom{kl}\mu\nu}\, E_k^{\mu}E_l^{\nu} -
           2N^{\mu}\,R^{\perp k}_{\phantom{\perp k}\mu\nu}\,E_k^{\nu}+
           2R^{\perp k}_{\phantom{\perp k}t\nu}\,E_k^{\nu}\right)
           \, dtd^3x\,,
                                                               &(4.26)\cr}
$$
where we simply write $dtd^3x$ for $dt\w dx^1\w dx^2\w dx^3$.
The first thing to observe is the purely polynomial dependence on all
field variables $N,N^{\mu},E_k^{\mu}$ and $\omega_{\alpha}^{ab}$. The reason
for this lies in the usage of the densitized variables (4.19-21). In a next
step we would like to calculate the Hamiltonian. For this we have to calculate
the momenta by differentiating with respect to the time derivatives of the
field variables. Obviously, $N,N^{\mu}$ and $E_k^{\mu}$ appear
undifferentiated.
As far as the connection variables are concerned, the first and second
terms in (4.26) only contain $\omega_{\mu}^{kl}$ and $\omega_{\mu}^{\perp l}$
and spatial derivatives thereof. The third term, however, contains
$\omega_{\mu}^{kl}$, $\omega_t^{kl}$, $\omega_t^{\perp k}$ and spatial
derivatives thereof, but also $\omega_{\nu}^{\perp k}$ with its time
derivative.
No other quantities appear with their time derivatives.

In calculating the Hamiltonian we would then have to go through a
constraint analysis along the standard algorithm as outlined in
(Dirac 1967) and spelled out in more detail in (Hanson, Regge and
Teitelboim 1976) or (Henneaux and Teitelboim 1992). We also recommend
(Gotay, Nester, Hinds 1978) and (Marmo, Mukunda and Samuel 1983) for
a more geometric approach. A good summary may be found in the lecture
by Andreas Wipf in this volume. Here we cannot go through the full
constraint analysis as it would present itself when started at this
point. We merely want to point out one of its important results that
may bee seen as being responsible for the strategy of complexification.
More details may be found in (Romano 1993).

In our case, the constraint analysis would start with the primary
constraints of vanishing
momenta for all field variables except $\omega_{\nu}^{\perp k}$, and then
continue with secondary constraints to ensure preservation of these during
Hamiltonian evolution. This system of constraints then turns out not to be
first class.
Solving the purely second-class constraints reintroduces a complicated
dependence on the field variables, which essentially brings us back to
the standard ADM treatment of canonical General Relativity.
This undesirable feature can be located as being due to the nondynamical
nature of the field components $\omega_{\mu}^{kl}$, i.e. their vanishing
momenta. The way complex General Relativity circumvents this problem is
via the self dual representation (the anti-self-dual representation would
just be as good), where
$$
\shalf\epsilon^{\perp k}_{\phantom{\perp k}lm}\, \omegap_{\alpha}^{lm}
      =i\, \omegap_{\alpha}^{\perp k}\,,
\eqno{(4.27)}
$$
so that the time derivatives of ${\omegap_{\mu}}^{lm}$ now appear through
those of ${\omegap_{\mu}}^{\perp k}$.
To see how much is achieved this way we have to explicitly 3+1-decompose
the self dual action of complex General Relativity. This is done in the
next section.

\beginsection{Chapter 5. 3+1-Decomposition of Self-Dual Action in
Ashtekar's Variables}

  From now on we only consider complex General Relativity with self
dual action. We omit the $(+)$ on $\omegap$ and simply write
$\omega$ for the self dual connection. Unless stated otherwise, all the
connections and curvatures are self-dual from now on.
We have
$$\eqalignno{
& \omega_{\alpha}^{\perp k}=
-i\shalf\epsilon^{\perp k}_{\phantom{\perp k}lm}\ \omega_{\alpha}^{lm}
&(5.1)\cr
& \omega_{\alpha}^{lm}=-i\,\epsilon^{lm}_{\phantom{lm}\perp p}\
\omega_{\alpha}^{\perp p}
&(5.2)\cr
& R^{\perp k}_{\phantom{\perp k}\alpha\beta}=-i\shalf\,
\epsilon^{\perp k}_{\phantom{\perp k}lm}\
R^{lm}_{\phantom{lm}\alpha\beta}\,,
&(5.3)\cr}
$$
which we use to convert all $R^{\perp k}_{\phantom{\perp k}\alpha\beta}$
into $R^{nm}_{\phantom{nm}\alpha\beta}$ and $\omega_{\alpha}^{\perp k}$
into
$\omega_{\alpha}^{nm}$ in the expression for the 3+1-decomposed self-dual
action. Moreover, we write antisymmetric complex self-dual matrices as
3-dimensional complex vectors, i.e., if $\lambda^{kl}$ is such a matrix,
we set:
$$\eqalign{
& \lambda^m=-i\shalf\epsilon^{\perp m}_{\phantom{\perp m}kl}\
            \lambda^{kl}\,,                                     \cr
& \lambda^{kl}=-i\epsilon^{kl}_{\phantom{kl}\perp m}\ \lambda^m
              =:-i\epsilon^{kl}_{\phantom{kl}m}\lambda^m \,,
  \cr}
\eqno{(5.4)}
$$
and abbreviate the 3-tuple $(\lambda^1,\lambda^2,\lambda^3)$ by
$\vec\lambda$.
We also use ``scalar-'' and ``vector-product'' notations:
$$\eqalignno{
& \sum_{k=1}^3\lambda^k\sigma^k=:\vec\lambda\cdot\vec\sigma &(5.5)\cr
&  i\epsilon^k_{\phantom{k}nm}\, \lambda^n\sigma^m:=
(\vec\lambda\times\vec\sigma)^k\,.
&(5.6)\cr}
$$
This allows us to use convenient identities like
$$\eqalign{
& \vec\lambda\cdot(\vec\sigma\times\vec\rho)=
   (\vec\lambda\times\vec\sigma)\cdot\vec\rho \,,               \cr
& \vec\lambda\times(\vec\sigma\times\vec\rho)=
\vec\rho\,(\vec\lambda\cdot\vec\sigma)-
  \vec\sigma\,(\vec\lambda\cdot\vec\rho)\,.                                 \cr
}
\eqno{(5.7)}
$$
Note that due to the factor of $i$ (which we included for later
convenience) in (5.6), the right hand side of the second formula (5.7)
carries the opposite sign to the familiar rule in vector calculus.
Using this notation, we have e.g. ($\lambda^k_{\,l}=\d_{ln}\lambda^{kn}$,
that is, indices are raised and lowered with the Kronecker delta)
$$\eqalignno{
& \lambda^k_{\,l}\sigma^l =(\vec\lambda\times\vec\sigma)^k
&(5.8) \cr
& [\lambda,\sigma]^{kl}:=(\lambda^k_{\,r}\sigma^{rl}-
                          \sigma^k_{\,r}\lambda^{rl})
                        =  \lambda^k\sigma^l-\lambda^l\sigma^k \,,
&(5.9) \cr
\hbox{hence}\quad
& -i\shalf\epsilon^{\perp m}_{\phantom{\perp m}kl}\
[\lambda,\sigma]^{kl} = (\vec\lambda\times\vec\sigma)^m  \,.   &(5.10)\cr}
$$
Using these formulae we can simply convert the curvature components
into a convenient 3-dimensional notation. The connection
becomes a 3-vector valued 1-form $\vec\omega$, and by
$\vec\omega\times\vec\omega$ we shall denote the vector valued 2-form
whose components are $i\epsilon^k_{\phantom{k}lm}\omega^l\w\omega^m$.
$$\eqalignno{
& R^{\perp k}_{\phantom{\perp k}\mu\nu}=
  (d\vec\omega+\vec\omega \times\vec\omega)^k_{\mu\nu}   &(5.11) \cr
& R^{kl}_{\phantom{kl}\mu\nu} = -i\epsilon^{kl}_{\phantom{kl}p}\
(d\vec\omega+\vec\omega\times\vec\omega)^p_{\mu\nu}      &(5.12) \cr
& R^{\perp k}_{\phantom{\perp k}t\nu} =
\partial_t\omega^k_{\nu}- \partial_{\nu}\omega_t^k-
2(\vec\omega_{\nu}\times\vec\omega_t)^k                  &(5.13) \cr}
$$

The next step is to bring these expressions into a 3-dimensional
{\it covariant} form, that is, into a form involving only geometric
quantities on the 3-dimensional manifold $\S$.
Note, for example, that the expression on the right hand side
of (5.11) differs by a relative factor of 2 from what would be the
curvature of a 3-dimensional connection $\vec\omega$; it cannot be, since
$\vec\omega_{\mu}$ does not even define a 3-dimensional
connection. But $2\vec\omega$ does! (connections form an affine
but not a linear space and cannot generally be added and multiplied with
numbers). An elementary way to see this is
the following (during this argument we reemploy the old notation
using $\omegap$): the spatial part of the self-dual projection for the
curvature 1-form is given by,
$
{\omegap_{\mu}}^{kl}=
\shalf({\omega_{\mu}}^{kl}-i\epsilon^{kl}_{\phantom{kl}\perp n}
{\omega_{\mu}}^{\perp n}).
$
${\omega_{\mu}}^{kl}$ is easily seen to form the
components of a spatial connection and ${\omega_{\mu}}^{\perp n}=K_{\mu}^n$,
form the components of the 3-dimensional tensor of extrinsic curvature
(as we shall independently show below in (7.11-12)).
Thus their sum (but not half of it) form a 3-dimensional connection.
A geometrically more satisfying argument,
which we suppress at this point, can be given using general bundle
language. It is however important to realize that the spacetime
structure group $SO(1,3)_{\C}$ is reduced to the spatial structure group
$SO(3)_{\C}$ which is that subgroup that stabilizes the normal $e_{\perp}$.
It is only with respect to this reduction that the components
${\omega_{\mu}}^{kl}$ form a 3-connection and ${\omega_{\mu}}^{\perp n}$ a
tensor.

We thus define
$$\eqalignno{
& \vec A_{\mu}:=2\vec\omega_{\mu}  &(5.14)  \cr
& \vec \Lambda:=-2\vec\omega_t     &(5.15)  \cr}
$$
(the minus sign in (5.15) being chosen for later convenience)
and have
$$\eqalignno{
& R^{\perp k}_{\phantom{\perp k}\mu\nu}=\shalf F^k_{\mu\nu}
&(5.16)\cr
& R^{kl}_{\phantom{kl}\mu\nu}= -i\shalf\epsilon^{kl}_{\phantom{kl}p}
F^p_{\mu\nu}\,,
&(5.17)\cr
\hbox{where}\quad & \vec F_{\mu\nu}=
(d\vec A+\shalf\vec A\times\vec A)_{\mu\nu}
&(5.18)\cr
\hbox{and}\quad & R^{\perp k}_{\phantom{\perp k}t\nu}=
\shalf(\partial_t A^k_{\nu}+D_{\nu}\Lambda^k)\,,
&(5.19)\cr
\hbox{where}\quad & D_{\nu}\vec\Lambda := \partial_{\nu}\vec\Lambda+
                                  \vec A_{\nu}\times \vec\Lambda\,.
&(5.20)\cr}
$$
We finally also abbreviate $(E_1^{\mu},E_2^{\mu},E_3^{\mu})=\vec {E^{\mu}}$
and regard the action as functional in the variables $\vec A_{\mu}$
$\vec {E^{\mu}}$, $N$ and $N^{\mu}$. These are Ashtekar's variables, or
connection variables, since $\vec A_{\mu}$ is a connection 1-form for a
$SO(3)_{\C}$-connection over the 3-dimensional manifold $\S$. Given the
relations above and the decomposition (4.26) of the action (4.25)
(now considered for the complex case), we can write the complex action
in the following form:
$$
S_{\C}=\int\left\{(\partial_t\vec A_{\mu})\cdot \vec{E^{\mu}}-\left[
\shalf N\,\vec F_{\mu\nu}\cdot(\vec{E^{\mu}}\times\vec{E^{\nu}})
+N^{\mu}\, \vec F_{\mu\nu}\cdot\vec{E^{\nu}}
+\vec\Lambda\cdot D_{\mu}\vec{E^{\mu}}\right]\right\}\,dtd^3x\,,
\eqno{(5.21)}
$$
where the last term has already been integrated by parts without keeping
the surface term. The action is now seen to be already in Hamiltonian form,
i.e., of the form
$$
             \int (\dot q p-\H(q,p))\,dt  \,,
\eqno{(5.22)}
$$
thus identifying the canonical conjugate pair of variables as
$$
\left(\vec A_{\mu}\,,\,\vec{E^{\mu}}\right)
\eqno{(5.23)}
$$
and the Hamiltonian
$$
\H = \int_{\S}\left\{
\shalf N\,\vec F_{\mu\nu}\cdot(\vec{E^{\mu}}\times \vec{E^{\nu}}) +
N^{\mu}\,\vec F_{\mu\nu}\cdot \vec{E^{\nu}} +
\vec\Lambda\cdot(D_{\nu}\vec{E^{\nu}})\right\}\,d^3x \,.
\eqno{(5.24)}
$$
In making this direct identification of the canonical variables from
the Hamiltonian variational principle we have jumped over a large
part of constraints analysis. Note that originally both, the connection and
the tetrad, were configuration-space variables and now turn out to be
canonical conjugates. This would have been seen in the constraint
analysis as an outcome of the explicit reduction of certain second-class
constraints (e.g. Romano 1993). On the other hand, we know of the equivalence
of the original action with the reformulated action (5.21), which can be
employed to circumvent a tedious piece of explicit reduction of
second-class constraints by starting directly from (5.21). This leaves us
with a purely first-class system, as will be shown in the following
chapter.

\beginsection{Chapter 6. Constraint Algebra}

If we vary the action with respect to the fields $N$, $N^{\mu}$ and
$\vec\Lambda$ we obtain equations in the canonical variables without time
derivatives, that is, we obtain the constraints:
$$\eqalignno{
& H := \shalf\vec F_{\mu\nu}\cdot (\vec{E^{\mu}}\times\vec{E^{\nu}}) = 0\,,
&(6.1)\cr
& H_{\mu} := \vec F_{\mu\nu}\cdot\vec{E^{\nu}} = 0\,,
&(6.2)\cr
& \vec H := D_{\mu}\vec{E^{\mu}} = 0\,.
&(6.3)\cr}
$$
The Hamiltonian is just the sum of smeared constraints with the
smearing functions $N$, $N^{\mu}$ and
$\vec\Lambda$:
$$
\H =
\int_{\S}\left\{ NH + N^{\mu}H_{\mu} + \vec\Lambda\cdot\vec H
\right\}\,d^3x =:\Sc[N] + \D[N^{\mu}] + \G[\vec\Lambda]\,,
\eqno{(6.4)}
$$
where the expressions on the right hand side are just abbreviations for the
integrals in their appearing order. The constraint equations on phase space
are properly interpreted as saying that for specified test-function spaces
for $N$, $N^{\mu}$ and $\vec\Lambda$  the expressions $\Sc(N)$, $\D(N^{\mu})$
and $\G(\vec\Lambda)$ have to vanish {\it for all} test fields $N$,
$N^{\mu}$ and $\vec\Lambda$. We call them the scalar-, diffeomorphism-, and
gauge-constraints respectively. Note that there are infinitely many of each
of them.

The actions of the scalar-, gauge-, and diffeomorphism-constraint
on the phase space are easily determined. Let $\{{},{}\}$ denote the
Poisson bracket with respect to the symplectic structure given to us
with the action (5.21) (in the standard way, namely by telling us that
$\vec A_{\mu}$ and $\vec{E^{\mu}}$ are Darboux coordinates on phase
space for the symplectic 2-form):
$$\eqalignno{
& \{\vec A_{\mu},\Sc[N]\}=N\,\vec {E^{\nu}}\times\vec F_{\mu\nu}
&(6.5)\cr
& \{\vec {E^{\mu}},\Sc[N]\}=D_{\nu}(N\,\vec{E^{\nu}}\times\vec{E^{\mu}})\,,
&(6.6)\cr
\hbox{and}\quad
& \{\vec A_{\mu},\G[\vec\Lambda]\}=-D_{\mu}\vec\Lambda
&(6.7)\cr
& \{\vec{E^{\mu}},\G[\vec\Lambda]\}=\vec\Lambda\times\vec{E^{\mu}}\,,
&(6.8)\cr
\hbox{and}\quad
& \{\vec A_{\mu},\D[N^{\mu}]\}=
L_{N}\vec A_{\mu}+\{\vec A_{\mu},\G[\vec A(N)]\}
&(6.9)\cr
& \{\vec{E^{\mu}},\D[N^{\mu}]\}=
L_{N}\vec {E^{\mu}}+\{\vec {E^{\mu}},\G[\vec A(N)]\}\,,
&(6.10)\cr}
$$
where $L_N$ denotes the Lie derivative with respect to the vector field
$N^{\mu}\partial_{\mu}$. Note that Lie derivatives of densities differ
from those
of the undensitized quantities by divergence terms in $N^{\mu}$. The Lie
derivative of the connection $\vec A_{\mu}$ is defined componentwise
by treating the connection as three one-forms. Note that this definition
makes only sense with respect to a global trivialization of the underlying
bundle (which we do have here), but not generally. We also set
$\vec A(N)=\vec A_{\mu}N^{\mu}$.

The first observation we make is, that due to the simple polynomial
form of the Hamiltonian we never had to invert the matrix
$\{e^{\mu}_k\}$ in order to calculate the Hamiltonian flow. This formally
gives rise to the possibility of generalized, degenerate solutions,
which have no counterpart in the metric formulation.

Formulae (6.7-10) show that $\G$ generates $SO(3)_{\C}$ frame rotations
and that $\D$ generates diffeomorphisms. It might appear that $\D$
also generates frame rotations, if one defines a ``pure'' diffeomorphism
by having only a Lie derivative on the right hand side of (6.9-10).
But such a definition does not make unambiguous sense since it is
gauge-dependent. In fact, the right hand side of (6.9-10) {\it is} gauge
independent. Nevertheless, with respect to a chosen trivialization,
such a definition of a pure diffeomorphism is possible and might
even be convenient for applications. In the present case, this amounts
to a redefinition of the diffeomorphism generator:
$$\eqalignno{
& H_{\mu}\mapsto H'_{\mu}:=H_{\mu}-\vec A_{\mu}\cdot \vec H
\,,\quad\hbox{or}\quad \D'[N^{\mu}]=\D[N^{\mu}] - \G[\vec A(N)]\,,
&(6.11)\cr
& \hbox{so that}\quad
  \{\vec A_{\mu}, \D'[N^{\mu}]\}=L_N\vec A_{\mu}
&(6.9')\cr
& \phantom{\hbox{so that}\quad}
  \{\vec{E^{\mu}},\D'[N^{\mu}]\}=L_N\vec{E^{\mu}}\,.
&(6.10')\cr}
$$

With a little bit of further work one can calculate the Poisson brackets
for all constraint functions:
$$\eqalignno{
& \{\G[\vec\Lambda],\G[\vec M]\}=\G[\vec\Lambda\times\vec M]
&(6.12)\cr
& \{\G[\vec\Lambda],\D[N^{\mu}]\}=0
&(6.13)\cr
\hbox{or}\quad
& \{\G[\vec\Lambda],\D'[N^{\mu}]\}=-\G[L_N\vec\Lambda]
&(6.13')\cr
& \{\G[\Lambda],\Sc(N)\}=0
&(6.14)\cr
& \{\D[N^{\mu}],\D[M^{\mu}]\}=\D[L_NM^{\mu}]-\G[\vec F(N,M)]
&(6.15)\cr
\hbox{or}\quad
& \{\D'[N^{\mu}],\D'[M^{\mu}]\}=\D'[L_NM^{\mu}]
&(6.15')\cr
& \{\D[N^{\mu}],\Sc[M]\}=\Sc[L_NM]+
\G[MN^{\mu}\vec F_{\mu\nu}\times\vec{E^{\nu}}]
&(6.16)\cr
\hbox{or}\quad
& \{\D'[N^{\mu}],\Sc[M]\}=\Sc[L_NM]
&(6.16')\cr
& \{\Sc[N],\Sc[M]\}=\D[K^{\mu}]=\D'[K^{\mu}]+ \G[\vec A(K)]
&(6.17)\cr
\hbox{where}\quad
& K^{\mu}:=
\vec{E^{\mu}}\cdot\vec{E^{\nu}}(M\partial_{\nu}N-N\partial_{\nu}M)\,.
&(6.18)\cr}
$$
First of all, we observe that the right sides are all proportional to
constraint functions with coefficients that generally depend on
phase space. Phase space dependencies appear in (6.15)(6.16)
and (6.17). In the first two cases these can be removed by using $\D'$
instead of $\D$ (equations (6.15') and (6.16')). So the only phase space
dependence appears in the Poisson bracket of two scalar constraints
in the bilinear combination $\vec{E^{\mu}}\cdot\vec{E^{\nu}}$
on the right hand side of (6.17). This prevents $\Sc,\D,\G$
from generating an (infinite dimensional) Lie algebra. For example,
we cannot calculate higher Poisson brackets using only the relations
(6.12-17), since we would need to know their Poisson-bracket with the
term (6.18), which cannot be inferred from the relations (6.12-17).
For this we would need to know the explicit functional forms of the
generators.

Relation (6.17) also tells us that the scalar constraints
do not close onto themselves, as it is the case with the $\G$ and the
$\D'$ generators individually. Moreover, the Poisson brackets for $\G$
with any
other generator is proportional to $\G$. We say that $\G$ generates an
ideal whereas the $\D'$ only generate a subalgebra (note the abuse of
language here).
For them to form an ideal too we should have the right hand side of
(6.16') to be proportional to a diffeomorphism-, not a scalar-constraint.
The geometric meaning of $\G$ generating an ideal is the
following: the phase space flow generated by $\D$ (or $\D'$) and $\Sc$
restricted to the constraint hypersurface $\G=0$ is tangential to this
surface and therefore leaves it invariant.
This allows us to reduce the constraints in steps, that is, to
{\it first} go to the $\G=0$ hypersurface and {\it then} reduce the action
generated by all other constraints. That such a separate phase space
reduction for the gauge generators exists is clear from the existence
of a formulation of the theory in terms of gauge invariant variables, like
the metric formulation. However, a similar procedure with the diffeomorphism
constraint is impossible, since its flow does not leave the $\Sc=0$
hypersurface invariant.
On the other hand, for example, if one were to determine functions in
phase space (or subsets thereof) which are invariant under the action of
the constraints, it
would be sufficient to look for functions annihilated by a generating
set of constraints. But due to the intertwinement of diffeomorphism-
with scalar-constraints, generating sets much smaller than the defining
set may actually be found.
For example, functions on the constraint
hypersurface $\Sc[N]=0=\D[N^{\mu}]$, with zero derivative along the
Hamiltonian
vector fields of $\Sc[N]$ for all $N$, have also zero derivative along the
Hamiltonian vector fields of $\D[K^{\mu}]$, due to (6.17-18).
But it is easy to see that every vector field can locally be written as sum
of vector fields of the form (6.18) so that $\Sc[N]$-invariant functions
are automatically $\D[N^{\mu}]$-invariant. Similar observations of this
type where first made in (Moncrief 1972) and (Moncrief and Teitelboim 1972).

\beginsection{Chapter 7. Reality Conditions and Geometric Interpretation}

So far we have reformulated complex General Relativity. But, of course,
ultimately we are interested in the ordinary real case. A central
question therefore is how to recover the real theory within the formalism
developed, that is, how to derive a real 3-metric from the complex
equations of motion for $\vec A_{\mu}$ and $\vec {E^{\mu}}$ with real
lapse and shift functions $N$ and $N^{\mu}$. In the real case it is also
easy to give some geometric interpretation to the variables $\vec A_{\mu}$
and $\vec {E^{\mu}}$. We shall briefly discuss these points in this
chapter.

As outlined in chapter 3, we are working on  complex tensor bundles with
reality structure. Real sections have real component functions with respect
to real bases and real bases are provided since we work on a real manifold.
Note that we only require the 3-metric and not the (densitized)
tetrad to be real. The former is derived from the latter according to:
$$\eqalignno{
& h^{\mu\nu}={1\over h}\sum_{k=1}^3 \vec {E^{\mu}}\cdot\vec{E^{\nu}}\,,
&(7.1)\cr
\hbox{so that}\quad
& h^{\mu\nu}=
\left[\hbox{det}\{\vec{E^{\mu}}\cdot\vec{E^{\nu}}\}\right]^{-\half}\,
\vec{E^{\mu}}\cdot\vec{E^{\nu}}\,.
&(7.2)\cr}
$$

Suppose now that $\vec{E^{\mu}}\cdot\vec{E^{\nu}}$ is real. The Hamiltonian
flow should not disturb this reality, that is
$$
\{\vec{E^{\mu}}\cdot\vec{E^{\nu}}\,,\, \Sc[N]+\D'[N^{\mu}]+\G[\vec\Lambda]\}
\,=\,\hbox{real}\,,
\eqno{(7.3)}
$$
for all $\vec\Lambda$ and all {\it real} $N$ and $N^{\mu}$.
The gauge generator Poisson-commutes with $\vec{E^{\mu}}\cdot\vec{E^{\nu}}$
(compare (6.8)) and therefore drops out of discussion. The diffeomorphisms
generator simply yields (compare (6.10')
$$
\{\vec{E^{\mu}}\cdot\vec{E^{\nu}}\,,\, \D'[N^{\mu}]\} =
L_N\left(\vec{E^{\mu}}\cdot\vec{E^{\nu}}\right)
\eqno{(7.4)}
$$
which is real for real $N^{\mu}$ (the components with respect to the real
basis $\partial_{\mu}$). So the diffeomorphism generator as well does not
disturb reality. We can thus focus on the scalar constraint. We have
$$\eqalign{
\{\vec{E^{\mu}}\cdot\vec{E^{\nu}}\,,\, \Sc[N]\}
=& \left[\vec{E^{\mu}}\cdot(\vec{E^{\nu}}\times\vec{E^{\sigma}})+
         \vec{E^{\nu}}\cdot(\vec{E^{\mu}}\times\vec{E^{\sigma}})\right]\,
         \partial_{\sigma} N                                            \cr
& + N\,(D_{\sigma}\vec{E^{\sigma}})\cdot (\vec{E^{\mu}}\times
\vec{E^{\nu}} + \vec{E^{\nu}}\times \vec{E^{\mu}})
    \cr
& + N\,\left[\vec{E^{\sigma}}\cdot (
            \vec{E^{\mu}}\times D_{\sigma}\vec{E^{\nu}}+
            \vec{E^{\nu}}\times D_{\sigma}\vec{E^{\mu}})\right]\,.     \cr}
\eqno{(7.5)}
$$
The first two expressions vanish identically, so that the resulting reality
condition is:
$$
\left(\vec{E^{\sigma}}\times D_{\sigma}\vec{E^{(\mu}}\right)
\cdot\vec{E^{\nu)}} \,=\,\hbox{real}\,.
\eqno{(7.6)}
$$
One may check that there are no further constraints, i.e., the Poisson
bracket of this expression with the Hamiltonian is real due to the reality
conditions already posed. So we have the result that the Hamiltonian flow
preserves reality provided the initial data are further constrained by the
reality of $h_{\mu\nu}$ and (7.6).

Now we turn to the geometric interpretation of the canonical
variables. We begin with the momenta $\vec{E^{\mu}}$. We write (7.1)
in the form
$$
\hbox{det}\{h_{\mu\nu}\}h^{\mu\nu}=\vec{E^{\mu}}\cdot\vec{E^{\nu}}
\eqno{(7.7)}
$$
and consider  e.g. the 3-3 component of this equation. The left hand
side is just
the subdeterminant $h_{11}h_{22}-h_{12}^2$ which for real $h_{\mu\nu}$
measures the area content
of the parallelogram in tangent-space spanned by the vectors
$\partial_1$ and $\partial_2$. The area content of an embedded 2-surface
(coordinatized by $\sigma$ and $\tau$):
$$
(\sigma, \tau)\mapsto x^{\mu}(\sigma, \tau)\,,
\eqno{(7.8)}
$$
is thus given by (no summation convention here):
$$
\int\sum_{\lambda ,\mu ,\nu}
\sqrt{\vec{E^{\lambda}}\cdot\vec{E^{\lambda}}}\ \varepsilon_{\lambda\mu\nu}\,
{\partial x^{\mu}\over\partial\tau}{\partial x^{\nu}\over\partial \sigma}\,
d\tau d\sigma\,,
\eqno{(7.9)}
$$
where the integral is taken over the surface.
$\varepsilon_{\lambda\mu\nu}$ denotes the 3-form density on $\S$ of
weight $-1$, whose components in {\it all} coordinate systems are $0$
if two indices coincide and $\pm 1$ if they form an even, respectively odd
permutation of $123$.
In that sense the variables $\vec{E^{\mu}}$ are geometrically connected
to areas rather than length.

Next we consider the variables $\vec A_{\mu}$. For this, we restore our
old notation using $\omega$ for the connection 1-form and $\omegap$ for
the self-dual part, and have
$$
A^k_{\mu}=2\omegap^k_{\mu}=
-i\shalf\epsilon^{\perp k}_{\phantom{\perp k}lm}(\omega_{\mu}^{lm}-
i\epsilon^{lm}_{\phantom{lm}\perp p}\omega_{\mu}^{\perp p})\,.
\eqno{(7.10)}
$$
If the equations of motion are satisfied we have $\omega=\Gamma$ (the
Levi-Civita connection). Let its spacetime covariant derivative be
${}^{(4)}\nabla$ and ${}^{(3)}\nabla$ its induced space covariant
derivative on $\S$ (compatible with $h_{\mu\nu}$). We recall the
definition of extrinsic curvature:
$$\eqalignno{
& {}^{(4)}\nabla_{e_k} e_l=:{}^{(3)}\nabla_{e_k} e_l + K_{kl}\,e_{\perp}
& (7.11)\cr
\hbox{and have}\quad
& \omega_{\mu}^{\perp k}=K_{\mu}^k=\delta^{kl}e^i_{\mu}K_{il}\,.
&(7.12)\cr}
$$
Inserting this into (7.10) yields
$$\eqalignno{
& A_{\mu}^k=\Gamma_{\mu}^k+K_{\mu}^k \,,
&(7.13)\cr
\hbox{where}\quad
& \Gamma_{\mu}^k:=
-i\shalf\epsilon^{\perp k}_{\phantom{\perp k}ln}\Gamma_{\mu}^{ln}\,.
&(7.14)\cr}
$$
The spatial components, $\Gamma_{\mu}^{ln}$, are of course just the
components for the Levi-Civita connection on $\S$ compatible with
$h_{\mu\nu}$.
Relation (7.13) clearly shows the non-triviality of the transformation
to the connection variables: the connection 1-form $\vec A$ involves
the metric and the extrinsic curvature and therefore both conjugate
sets of variables of the metric formulation. We also note that (7.13)
gives rise to an alternative way of expressing a reality condition
which proves useful in a variety of applications, in particular when
compared with the standard metric formulation.
The reality of $K_{\mu}^k$ and $\Gamma_{\mu}^{ln}$ (hence $\Gamma_{\mu}^k$
is purely imaginary) implies (a bar denotes complex conjugation)
$$
A^k_{\mu}-{\bar A}^k_{\mu}=2\Gamma_{\mu}^k\,.
\eqno{(7.15)}
$$

Finally, we point out some further interesting decomposition of the
connection 1-form $\vec A$. First of all, using the dual basis
$\{\theta^k\}$ of $\{e_k\}$ on $\S$ (as introduced before equation
(4.2)), we have $A^k=A^k_{\mu}dx^{\mu}=A^k_l\theta^l$. The latin
indices are now raised and lowered using $\delta_{kl}$, ie.,
$A_{kl}=\delta_{kn}A^n_l$. Correspondingly, (7.13) reads:
$A_{kl}=\Gamma_{kl}+K_{kl}$. Due to the symmetry of $K_{kl}$
(which follows from (7.11) and the vanishing torsions of
${}^{(4)}\nabla$ and ${}^{(3)}\nabla$) the antisymmetric part of $A_{kl}$
just involves $\Gamma_{kl}$. Defining the quantity
$\Gamma_l:={i\over 2}\epsilon_l^{\phantom{l}nm}\Gamma_{nm}$,
which is real, we have
$$
A_{kl}=A_{(kl)}-i\epsilon_{kl}^{\phantom{kl}n}\Gamma_n
      =\Gamma_{(kl)}+K_{kl} -i\epsilon_{kl}^{\phantom{kl}n}\Gamma_n\,.
\eqno{(7.16)}
$$
We further set
$$
A:=\delta^{kl}A_{kl}=\delta^{kl}\Gamma_{(kl)}+\delta^{kl}K_{kl}
  =:\Gamma+K\,,
\eqno{(7.17)}
$$
where $K$, the trace of the extrinsic curvature, is zero, iff $\S$ is a
maximal hypersurface (i.e. a stationary point of the area functional).
Using the first Cartan structure equation for the basis $\{\theta^k\}$
and the torsion-free Levi-Civita connection on $\S$, one easily shows
the following identities:
$$\eqalignno{
& \Gamma_k\theta^k = \shalf * \delta_{kl}(\theta^k\w * d\theta^l)\,,
                 &(7.18)\cr
& \Gamma = \shalf i * \delta_{kl}(\theta^k\w d\theta^l)\,,
                 &(7.19)\cr}
$$
where $*$ is now the 3-dimensional Hodge-duality map on $\S$
(defined analogously to (1.20)).
$\Gamma_k\theta^k$ and $\Gamma$ are a {\it frame dependent} 1-form and a
{\it frame dependent} function respectively (recall the
transformation law for connections (1.19)) and have been employed in
(Nester 1989) to fix the gauge freedom in the choice of frames
by imposing the gauge conditions:
$$\eqalignno{
                   & d(\Gamma_k\theta^k)= 0\,,  &(7.20)\cr
                   & d \Gamma           = 0\,.  &(7.21)\cr}
$$
If we insert the Levi-Civita connection with respect to an arbitrary
rotated frame by using formula (1.19) to parameterize it by the
rotation matrix $R$, these conditions can be seen
to form a system of non-linear elliptic differential equations for $R$
which under certain circumstances fixes it up to a constant rotation.
Since we assume $\S$ to be connected, (7.20) is clearly equivalent
to $\Gamma=const.$. If we make the additional assumption that $\S$ has
trivial first DeRahm Cohomology (equivalently, the space of harmonic
1-forms is zero dimensional. A sufficient condition for this is that
$\S$ has finite fundamental group.), then all closed 1-forms are exact
so that $\Gamma_k=e_k(\rho)$ for some globally defined function $\rho$
on $\S$. In this integrated form the gauge conditions have been shown
to be equivalent to a massive 2-spinor equation (the mass being given
by the value $\Gamma$), which is {\it linear} and elliptic (Dimakis and
M\"uller-Hoissen 1989). For asymptotically flat $\S$, the fall-off
conditions require $\Gamma=0$, in which case existence and uniqueness
of the massless spinor equation has been demonstrated, up to possible
isolated singularities, in (Ashtekar and Horowitz 1983). For a
neighborhood of flat 3-space, existence and uniqueness can be shown
directly (Nester 1989, also Nester 1992).

Another interesting question that plays a significant r\^ole in the
investigation of the initial data problem is, what spacetimes can be
evolved from maximal (i.e. $K=0$) hypersurfaces $\S$. It has been
shown (Bartnik 1984) that all spacetimes sufficiently close (in a sense
specified by Bartnik) to Minkowski space allow such maximal slicings.
Conversely, it is also known that there are even topological restrictions
on $\S$ in order to allow for any maximal initial data (Witt 1987).
The manifolds ruled out are those which do not allow for metrics
with everywhere positive Ricci scalar (as is directly seen from the
Hamiltonian constraint in the standard metric representation). They all
have infinite fundamental group.

At the end of the next section we will show how the special gauge
conditions (7.20-21) can be used to prove a restricted version of
the positive mass theorem which applies to maximal $\S$ (i.e.
spacetimes admitting such hypersurfaces).

\beginsection{Chapter 8. Surface Integrals}

If the spatial manifold $\S$ is not closed one has to specify fall-off
conditions for the field $\vec A_{\mu}$ and $\vec{E^{\mu}}$. These must
be chosen wide enough to accommodate physically interesting solutions, with,
say, non-vanishing energy, momentum and angular momentum etc.
at spatial infinity (we call these quantities the Poincar\'e charges).
In this case the Hamiltonian must be amended by surface integrals in order
to make it a continuously differentiable function on phase space. Note that
the point is not that otherwise we would get the wrong equations of
motion, but rather that we would not get any equations at all. Also, the
variational principle need a careful restatement in order to ensure
stationary points in the right class of functions. However, in order
to arrive at the right Hamiltonian it is not necessary to restart
from a modified variational principle to calculate the modified
Hamiltonian by keeping all non-vanishing boundary integrals. A much simpler
strategy is to first calculate a formal Hamiltonian ``brute force''
without keeping track of
any surface integrals, and then amend it ``by hand'' with the right surface
terms that make it a continuously differentiable function. For the metric
formulation of General Relativity this is explained and carried through
in (Regge and Teitelboim 1979) and in more detail in
(Beig and \'O Murchadha 1987), which also corrects a mistake in the first
reference concerning the generator of boosts. Using the connection
variables, an analysis similar to the one by Beig and \'O Murchadha was
performed in (Gl\"o\ss ner 1992) and also in (Thiemann 1993).
An alternative approach not within the canonical framework is
presented in (Ashtekar and Romano 1993).
Here we cannot give the full details of the general procedure but only
want to outline the underlying ideas.

We already have the (formal) Hamiltonian without the surface terms.
Let us restrict to $\vec\Lambda=0=N^{\mu}$ so that the (formal)
Hamiltonian equals the scalar constraint:
$$
\Sc[N]=\shalf\int_{\S} N\vec F_{\mu\nu}\cdot(\vec{E^{\mu}}
\times\vec{E^{\nu}})  \,d^3x \,.
\eqno{(8.1)}
$$
This expression is clearly differentiable with respect to $\vec{E^{\mu}}$
(since it enters undifferentiated), but if we try to differentiate it with
respect to $\vec A_{\mu}$ we encounter the surface integral
$$
-\int_{S_{\infty}}
N\,\vec A_{\mu}\cdot (\vec{E^{\mu}}\times\vec{E^{\nu}})\, d\sigma_{\nu}\,,
\eqno{(8.2)}
$$
where $\Si$ is the 2-sphere at spatial infinity. The integral over it is
understood in the usual way: take the integral over 2-spheres in the finite,
such that the result depends on the radius, and {\it then} take the limit
as the radius approaches infinity. Here we assume of course that it be finite.
Now, if this surface term is not zero, the functional $\Sc[N]$ is not
differentiable with respect to $\vec A_{\mu}$. We restore differentiability
by simply subtracting (8.2) from (8.1) and obtain:
$$
\H_s[N]:=
\shalf\int_{\S} N\vec F_{\mu\nu}\cdot(\vec{E^{\mu}}
\times\vec{E^{\nu}})\,d^3x
+\int_{S_{\infty}} N\,\vec A_{\mu}\cdot
(\vec{E^{\mu}}\times\vec{E^{\nu}})\, d\sigma_{\nu}\,,
\eqno{(8.3)}
$$
which we call the scalar Hamiltonian.
The constraints are still given by (6.1), or, in smeared form, by the
requirement that the first term in (8.3) vanishes. On the constraint
surface the scalar Hamiltonian is thus non-vanishing and assumes the
values of minus the surface integral (8.2). In the non-closed case
the Hamiltonian is not the sum of the constraints and does not vanish
on the constraint surface. Rather, it defines non-trivial functions
which descend to functions on the reduced phase space, i.e., to
non-trivial observables. For the case at hand one expects this quantity to
be the total energy of the isolated system described by asymptotically
flat data, and one may wonder whether the quantity above is actually real.
This can be quite easily demonstrated. But to do so, we have to first make
slightly more precise what we define as asymptotically flat data.
Asymptotically we require $N(r\rightarrow\infty)\rightarrow const.$
(so we leave out boosts from the discussion given here since for them
$N\propto x^{\mu}$ for large $r$.)
$$\eqalignno{
& E^{\mu}_k = \delta^{\mu}_k + {\alpha^{\mu}_k(\theta ,\varphi)\over r}
                             +  O\left({1\over r}\right)\,,
& (8.4)\cr
& A_{\mu}^k = {\beta_{\mu}^k(\theta ,\varphi)\over r^2}
              + O\left({1\over r^2}\right) \,,
&(8.5)\cr}
$$
where $\alpha^{\mu}_k$ and $\beta_{\mu}^k$ are functions on the 2-sphere
which also obey the conditions to be of even-, respectively odd parity under
the antipodal map of the sphere. We refer to (Beig, \'O Murchadha 1987)
for a discussion of these conditions in the metric case.
$O({1\over r^n})$ stands for terms
with asymptotic radial fall-off faster than $\propto 1/r^n$.
Using formulae (7.13-14)
we can write the surface contribution to (8.3) ``on shell'' (i.e., the
equations of motion being satisfied) in the form
$$
\int_{S_{\infty}} N\, E^{\mu}_kE^{\nu}_l
(\Gamma_{\nu}^{kl}-i\epsilon^{kl}_{\phantom{kl}\perp p}K_{\nu}^p)\,
d\sigma_{\mu}\,.
\eqno{(8.6)}
$$
In a real basis the first term is real and the second purely imaginary.
But the second vanishes due to the symmetry of $K_{kl}$ and the asymptotic
conditions (8.4). The first term can be shown to be just one-half times
the ADM energy expression (which means that the properly normalized action
is twice ours in units where $16\pi G/c^4=1$, as was already remarked in
the discussion following equation (3.21)).

Let us also have a look at the diffeomorphism constraint. In the same way
as above one immediately sees that differentiability with respect to
$\vec A_{\mu}$ is obstructed by the surface integral
$$
-2\int_{S_{\infty}} N^{[\mu}\vec{E^{\nu]}}\cdot
\vec A_{\mu}\,d\sigma_{\nu}\,,
\eqno{(8.7)}
$$
which we must subtract to obtain the differentiable
diffeomorphism-Hamiltonian
$$
\H_d[N^{\mu}]:= \int_{\S} N^{\mu}\vec F_{\mu\nu}\vec{E^{\nu}}\, d^3x+
2\int_{S_{\infty}} N^{[\mu}\vec{E^{\nu]}}\cdot\vec A_{\mu}\,d\sigma_{\nu}\,.
\eqno{(8.8)}
$$
Note that due to the parity conditions on $\beta_{\mu}^k$ the surface
integral is also finite for asymptotic rotations, where
$N^{\mu}\propto \varepsilon_{\mu\nu\lambda}n^{\nu}x^{\lambda}$
with constant $n^{\mu}$. ``On shell'', (8.8) becomes
$$
\H_d[N^{\mu}]= 2 \int_{S_{\infty}} (K_{\nu}^k-
i\epsilon^{\perp k}_{\phantom{\perp k}lm}
\Gamma_{\nu}^{lm})N^{[\mu}E_k^{\nu]}\,d\sigma_{\mu}\,.
\eqno{(8.9)}
$$
Again it can be shown that the imaginary part vanishes. The real part is
$$
-\int_{S_{\infty}}
N^{\nu}(K_{\nu}^{\mu}-\delta_{\nu}^{\mu}K^{\lambda}_{\lambda})
\, d\sigma_{\mu}
\eqno{(8.10)}
$$
which is again one-half times the ADM-expression for momentum and angular
momentum.

Let us finally mention the gauge constraint. Here the only surface
contribution
comes from the requirement of differentiability with respect
to $\vec{E^{\mu}}$ and is given by:
$$
 \int_{S_{\infty}} \vec\Lambda\cdot \vec{E^{\mu}}\,d\sigma_{\mu}\,.
\eqno{(8.11)}
$$
But here one probably does not want to have an extra charge associated
with the asymptotic gauge (frame) rotations for which the following
fall-off is sufficient ($n_{\mu}$ is the normal to the asymptotic 2-sphere)
$$
\Lambda^k \delta_k^{\mu}n_{\mu}\propto O\left({1\over r^2}\right)
\eqno{(8.12)}
$$
Essentially the same condition follows, for example, from the requirement
of invariance of angular momentum under gauge transformations, i.e.,
frame rotations. Since the Poincar\'e charges are given by integrals
with non manifestly gauge-invariant integrands (due to the explicit
apparence of $\vec A_{\mu}$) one generally has to explicitly check
compatibility of the chosen fall-off conditions with the requirement
of gauge invariance.

We remark that it is indeed possible to write down surface
integrals that make the Hamiltonian  continuously differentiable and
where the lapse and shift function spaces are chosen wide enough to
accommodate for all asymptotic Poincar\'e charges. As expected, it
turns out that the boost generator with asymptotic behaviour
$N\propto x^{\mu}$ requires the most subtle discussion to establish
finiteness and differentiability. This is as in the metric case --
compare (Beig and \'O Murchadha) --, and has in fact a similar
solution (Gl\"o\ss ner 1992). There is thus an important difference
between the closed and open case. In the former,
all of dynamics is entirely generated by constraints, whereas in the
latter the constraints generate only asymptotically (at spatial infinity)
trivial changes, in contrast to the Hamiltonian, which also generates
dynamics on quantities at infinity. No effective changes remain in the
closed case after having divided out those generated by the constraints.
In contrast, in the asymptotically flat case, there are still residual
motions which are generated by the Poincar\'e group.

We want to end this section by demonstrating positivity of the
surface integral in (8.3) for the special cases where $\S$ is
maximal (i.e. K=0). In doing this we essentially follow
(Nester 1994) with some adaptions in notation. The idea is to
rewrite the scalar Hamiltonian (8.3) into a form in which positivity
properties can be deduced,
possibly by suitably choosing the lapse density $N$ in the interior of
$\S$. But since on the constraint set the surface integral equals in
value the scalar Hamiltonian and is independent of the lapse density
in the interior of $\S$, this will also show positivity of the
surface integral. We start by simply rewriting the
scalar Hamiltonian (8.3) in terms of forms by using the dual basis
$\{\theta^k\}$, equation (4.19) and
$\sqrt{h}d^3x=\theta^1\w\theta^2\w\theta^3$:
$$
\H_s[N]=i\int_{\S} N\sqrt{h}\delta_{kl}F^k\w \theta^l
       - i\int_{S_{\infty}}N\sqrt{h}\delta_{kl}A^k\w \theta^l\,.
\eqno{(8.13)}
$$
Note that $N\sqrt{h}$ is a scalar function and equal to the lapse before
the redefinition (4.21) was made. Converting the surface integral into a
volume integral and expanding the forms in terms of $\theta^k$s yields
$$\eqalign{
\H_s[N]=
&  \shalf\int_{\S} N\sqrt{h}\left(A^{kl}A_{lk}-A^2\right)\,\sqrt{h}d^3x
\cr
& -\int_{\S} N\sqrt{h}\left(A^{kl}\Gamma_{lk}-A\Gamma\right)\,\sqrt{h}d^3x
\cr
& -\int_{\S} \epsilon^{klm}e_k(N\sqrt{h})\,iA_{lm}\,\sqrt{h}d^3x
\cr}
\eqno{(8.14)}
$$
using the notation defined at the end of the previous section.
We now impose the reality condition (7.15), just transcribed to the
components $A_{kl}$ with respect to the real basis $\{\theta^k\}$,
and decompose $A_{kl}$ into symmetric and antisymmetric part according
to (7.16) and obtain the real expression:
$$\eqalign{
\H_s[N]
=& \shalf\int_{\S} N\sqrt{h}\left({\bar A}^{(kl)}A_{(kl)}-
{\bar A}A\right) \,\sqrt{h}d^3x                         \cr
+& \int_{\S} N\sqrt{h} \Gamma_k\Gamma^k\,\sqrt{h}d^3x   \cr
-& 2 \int_{\S} e_k(N\sqrt{h})\Gamma^k\,\sqrt{h}d^3x\,.  \cr}
\eqno{(8.15)}
$$
Finally, we impose the special orthonormal frame gauge (7.20-21) in
integrated form: $\Gamma=0$ and $\Gamma_k=2e_k(\ln(\lambda))$, where
the way of writing the gradient is just chosen for convenience
with an everywhere positive and asymptotically constant function
$\lambda$. We then arrive at
$$\eqalign{
H_s[N]
=& \shalf\int_{\S} N\sqrt{h}\left({\bar A}^{(kl)}A_{(kl)}-K^2\right)
\,\sqrt{h}d^3x    \cr
& -4\int_{\S} \delta^{kl}e_k\left({N\sqrt{h}\over\lambda}\right)e_l(\lambda)
\,\sqrt{h}d^3x\,. \cr}
\eqno{(8.16)}
$$
The second integral can be made to vanish with the choice
$N\sqrt{h}=\lambda$, in which case the Hamiltonian simply reads
$$\H_s[\lambda]
=\shalf\int_{\S} \lambda\left({\bar A}^{(kl)}A_{(kl)}-K^2\right)
\,\sqrt{h}d^3x    \,.
\eqno{(8.17)}
$$
For maximal hypersurfaces $\S$ one has $K=0$ and the integrand is
non-negative and zero only for $A_{(kl)}=0$, which implies (real and
imaginary part vanish separately) $\Gamma_{(kl)}=0$ and $K_{kl}=0$.
Geometrically, the latter equation just says that $\S$ is totally
geodesic.

\vfill\eject

\beginsection{References}

{\parindent=0pt

{\bf Ashtekar, A.} (1991).  Lectures on Non-Perturbative Canonical
Gravity. Advanced Series in Astrophysics and Cosmology, Vol. 6,
World Scientific, Singapore - New Jersey - London - Hong Kong.

{\bf Ashtekar, A., Horowitz, G.} (1983). Jour. Math. Phys., {\bf 25},
                                         1473.

{\bf Ashtekar, A., Romano, J.D.} (1992). Class. Quant. Grav., {\bf 9},
                                         1069.

{\bf Bartnik, R.} (1984). The Existence of Maximal Surfaces in
                          Asymptotically Flat Spacetimes. In:
Asymptotic Behaviour of Mass and Spacetime Geometry, F. Flaherty (ed.),
Lecture Notes in Physics, 202, pp 57-60, Springer Verlag.

{\bf Beig, R., \'O Murchadha, N.} (1987). Ann. Phys. (N.Y.), {\bf 174}, 463.

{\bf Br\"ugmann, B.} (1993).
Bibliography of Publications related to Classical and Quantum Gravity
in terms of Ashtekar Variables. Los Alamos bulletin-board preprint
gr-qc/9303015.

{\bf Dimakis, A., M\"uller-Hoissen, F.} (1989). Phys. Lett. A,
                                        {\bf 142}, 73.

{\bf Dirac, P.A.M.} (1967).
Lectures on Quantum Mechanics. Yeshiva University, New York:
Academic Press.

{\bf Gl\"o\ss ner, P.} (1992). \"Uber die Phasenraumconstraints der
Algemeinen Relativit\"ats\-theorie in metrischer Formulierung und in der
Zusammenhangsdarstellung Ashtekars. Diploma Thesis, Freiburg University,
December 1992.

{\bf Gotay, M.J., Nester, J.M., Hinds, G.} (1978).
Jour. Math. Phys {\bf 19}, 2388.

{\bf Hanson, A., Regge, T., Teitelboim, C.} (1976).
Constrained Hamiltonian Systems. Rome: Acad. Naz. dei Lincei.

{\bf Henneaux, M., Teitelboim, C.} (1992).
Quantizations of Gauge Systems. Princeton University Press,
Princeton, New Jersey.

{\bf Marmo, G., Mukunda, N., Samuel, J.} (1983).
Revista del Nuovo Cimento
{\bf 6}, 1.

{\bf Moncrief, V.} (1972). Phys. Rev. D, {\bf 5}, 277.

{\bf Moncrief, V., Teitelboim, C.} (1972). Phys. Rev. D, {\bf 6}, 966.

{\bf Nester, J.} (1989). Jour. Math. Phys., {\bf 30}, 624.

{\bf Nester, J.} (1992). Jour. Math. Phys., {\bf 33}, 910.

{\bf Nester, J.} (1994). Los Alamos bulletin-board preprint gr-qc/9401004.

{\bf Regge, T., Teitelboim, C.} (1979). Ann. Phys. (N.Y.), {\bf 88}, 286.

{\bf Romano, J.D.} (1993). Gen. Rel. Grav. {\bf 25}, 759.

{\bf Thiemann, T.} (1993). Los Alamos bulletin-board preprint gr-qc/9310039.

{\bf Witt, D.} (1987). Toplogical Obstructions to Maximal Slices.
                       Santa Barbara Preprint, UCSB-1987.

}

\vfill\eject

\end